\documentclass[conference]{IEEEtran}
\IEEEoverridecommandlockouts

\usepackage{cite}
\usepackage{amsmath,amssymb,amsfonts}
\usepackage{algorithmic}
\usepackage{graphicx}
\usepackage{textcomp}
\usepackage{xcolor}
\usepackage[caption=false,font=footnotesize]{subfig}
\def\BibTeX{{\rm B\kern-.05em{\sc i\kern-.025em b}\kern-.08em
    T\kern-.1667em\lower.7ex\hbox{E}\kern-.125emX}}
\begin{document}

\title{Information in Polarization, Energy from Optical Power: Stokes-Orthogonal Inter-Satellite Links}

\author{
\IEEEauthorblockN{
Meysam~Ghanbari\IEEEauthorrefmark{1},
Mohammad~Taghi~Dabiri\IEEEauthorrefmark{2},
Mazen~O.~Hasna\IEEEauthorrefmark{3},
and~Khalid~A.~Qaraqe\IEEEauthorrefmark{1}
}

\IEEEauthorblockA{
\IEEEauthorrefmark{1}College of Science and Engineering, Hamad Bin Khalifa University, Doha, Qatar.\\
}

\IEEEauthorblockA{
\IEEEauthorrefmark{2}Electrical Engineering Division, Department of Engineering, University of Cambridge, Cambridge, UK.\\
}

\IEEEauthorblockA{
\IEEEauthorrefmark{3}Department of Electrical Engineering, Qatar University, Doha, Qatar.\\
Email: megh89467@hbku.edu.qa
}
}

\maketitle

\begin{abstract}
Joint information and energy transfer over inter-satellite free-space optical links is commonly based on explicit resource splitting, which couples communication reliability and energy delivery. This paper proposes a constant-total-power Stokes-orthogonal architecture that conveys information through polarization while preserving the received optical power for photovoltaic conversion. Complementary polarization branches support balanced information detection through their ac components, while their dc components drive matched multi-junction photonic power converters. The developed framework jointly characterizes pointing-impaired propagation, bit-error rate, outage, and nonlinear harvested power. The analysis establishes that the two information symbols only exchange the branch powers, leaving the unordered input-power pair unchanged; hence, the total harvested energy remains symbol invariant even under nonlinear conversion, without an explicit information-energy power split or linear-efficiency approximation. Numerical results confirm the analytical model, with the closed-form BER approximation remaining within $0.43\%$ of the exact evaluation over the examined pointing conditions. Compared with ideal lossless power splitting, the proposed architecture avoids the conventional BER--harvested-power tradeoff and provides an analytical baseline for power-neutral joint information and energy transfer in inter-satellite optical links.
\end{abstract}


\section{Introduction}

Optical wireless crosslinks are central to high-capacity spacecraft networks because they combine narrow beams, large carrier bandwidth, and low terminal size, weight, and power. Their practical design nevertheless depends on acquisition, tracking, beam divergence, terminal apertures, and residual pointing dynamics~\cite{ref1}. Compact terminal studies and mission designs, including CLICK, LaserCube, and DLR CubeSat terminals, document near-infrared links over distances extending from tens to thousands of kilometers and illustrate the stringent coupling between optical design and pointing performance~\cite{ref2,ref3,ref4}.

For a vacuum inter-satellite path, atmospheric absorption, scattering, and turbulence are absent, but Gaussian diffraction and finite-aperture collection remain fundamental. A lateral offset between the received beam and the aperture can materially reduce the collected fraction. Gaussian-beam propagation provides the diffraction scale~\cite{ref5}; exact circular-aperture collection can be expressed through the Marcum $Q$-function~\cite{ref6}; and the equivalent-beam pointing model of Farid and Hranilovic yields a tractable bounded power-law channel distribution~\cite{ref7}. Retaining both the exact aperture law and the equivalent-beam approximation is important: the former provides the geometric reference, while the latter supports closed-form averaging.

Simultaneous lightwave information and power transfer (SLIPT) has been investigated through transmitter adaptation, time or power allocation, receiver operating-point control, and photovoltaic reception \cite{ref8,ref9,10}. These approaches generally introduce a coupling between information performance and harvested energy because the two functions share optical, electrical, or receiver resources. Simultaneous dc power extraction and ac data reception from photovoltaic devices have also been demonstrated experimentally \cite{ref11}, while high-efficiency near-infrared multi-junction photonic power converters (PPCs) provide practical benchmarks for optical power conversion \cite{ref12}. More recently, nonlinear energy-harvesting models and achievable-rate analyses have been developed for multi-junction photovoltaic SLIPT receivers \cite{14}.

Polarization provides a distinct signaling degree of freedom. Jones and Stokes representations permit orthogonal polarization states to convey information without changing the total optical power \cite{ref13}. Polarization modulation has been investigated for free-space optical (FSO) communication \cite{16}, while polarization-division multiplexing has also been considered for inter-satellite optical links \cite{17}. These studies use polarization primarily as an information-bearing or multiplexing dimension, whereas established SLIPT architectures predominantly encode information through optical intensity or introduce a transmission, reception, or operating-point allocation between information and energy functions.

The gap addressed here is therefore more specific. A binary symbol is conveyed by changing the Stokes state while preserving the total launched optical power. After polarization separation, the two symbols only exchange two complementary branch powers. Their difference carries the information, whereas matched nonlinear energy converters receive the same unordered pair of optical inputs for either symbol. Consequently, balanced information recovery and symbol-invariant nonlinear energy harvesting can coexist without an explicit optical information–energy allocation factor.
The main contributions of this work are summarized as follows:
\begin{itemize}

    \item A constant-total-power Stokes-orthogonal architecture is proposed for inter-satellite SLIPT, enabling differential ac information recovery and dc energy harvesting through a PBS-separated dual-PPC receiver.

    \item A pointing-aware analytical framework is developed to characterize the conditional and average BER, SNR outage, and harvested power under exact and equivalent-beam aperture models.

    \item Symbol-invariant nonlinear energy harvesting is established for matched converter branches, with extensions to different energy-extraction timescales and branch mismatch.

    \item An idealized Lambert-$W$ PPC model is used to quantify harvested power and benchmark the proposed architecture against ideal lossless power splitting.
    
\end{itemize}

\section{System Model}

Fig.~1 illustrates the considered inter-satellite FSO architecture for simultaneous information transfer and optical energy harvesting. The transmitting satellite applies constant-power Stokes/polarization encoding before Gaussian-beam propagation. Following pointing displacement and finite-aperture collection, the receiving terminal compensates the polarization transformation and separates the field through a polarizing beam splitter (PBS). The resulting horizontal and vertical optical powers illuminate matched PPC branches; ac extraction supplies a balanced information detector, while the dc outputs are combined for energy harvesting.

\subsection{Scope and Physical Assumptions}

Consider a line-of-sight optical crosslink between two spacecraft separated by range $L$. The transmitter launches a fundamental Gaussian mode of constant optical power $P_t$ and wavelength $\lambda$. Vacuum propagation introduces diffraction, finite-aperture collection, and pointing displacement. Terminal insertion losses are collected into measured or declared efficiencies rather than hidden inside an additional ``channel loss,'' thereby preventing geometric loss from being counted again after circular-aperture integration.
The residual boresight error is sufficiently small that the beam-centroid displacement at the receiver aperture plane is $L\theta$ to first order.
Pointing is quasi-static during a symbol and independent between analytical fading draws.
A coded system may instead experience block pointing; the conditional expressions remain valid, while only the temporal averaging rule changes.

\subsection{Constant-Power Stokes-Orthogonal Alphabet}

Let $s\in\{+1,-1\}$ be equiprobable. In a tracked horizontal/vertical (H/V) basis, the Jones vector and corresponding Stokes vector are
\begin{equation}
\mathbf{e}_s
=
\sqrt{\frac{P_t}{2}}
\begin{bmatrix}
\sqrt{1+s}\\
\sqrt{1-s}
\end{bmatrix},
\qquad
\mathbf{S}_s
=
\begin{bmatrix}
P_t & sP_t & 0 & 0
\end{bmatrix}^{\mathrm{T}}.
\label{eq:stokes_alphabet}
\end{equation}

Thus, $S_0=P_t$ for every symbol while $S_1=\pm S_0$. Under the standard Stokes convention~\cite{ref13},
\begin{equation}
\begin{aligned}
S_0 &= |E_H|^2 + |E_V|^2, 
&
S_1 &= |E_H|^2 - |E_V|^2,
\\
S_2 &= 2\operatorname{Re}\{E_H E_V^{*}\},
&
S_3 &= -2\operatorname{Im}\{E_H E_V^{*}\}.
\end{aligned}
\label{eq:stokes_parameters}
\end{equation}

The two symbols are orthogonal, have equal energy, and require neither optical amplitude blanking nor a receiver-side information/energy split ratio.
A deterministic lossless polarization transformation $\mathbf{U}\in\mathbb{C}^{2\times2}$ is unitary. A tracked compensator $\mathbf{U}^{H}$ therefore restores the H/V basis without changing total power. Residual polarization imperfection is represented by a single contrast parameter $\kappa$; it is not treated as an additional random fading process. Time-varying uncompensated polarization, polarization-dependent loss, and branch mismatch lie outside the symmetric baseline and are identified when their consequences are discussed.

\begin{figure}[t]
    \centering
    \includegraphics[width=\columnwidth]{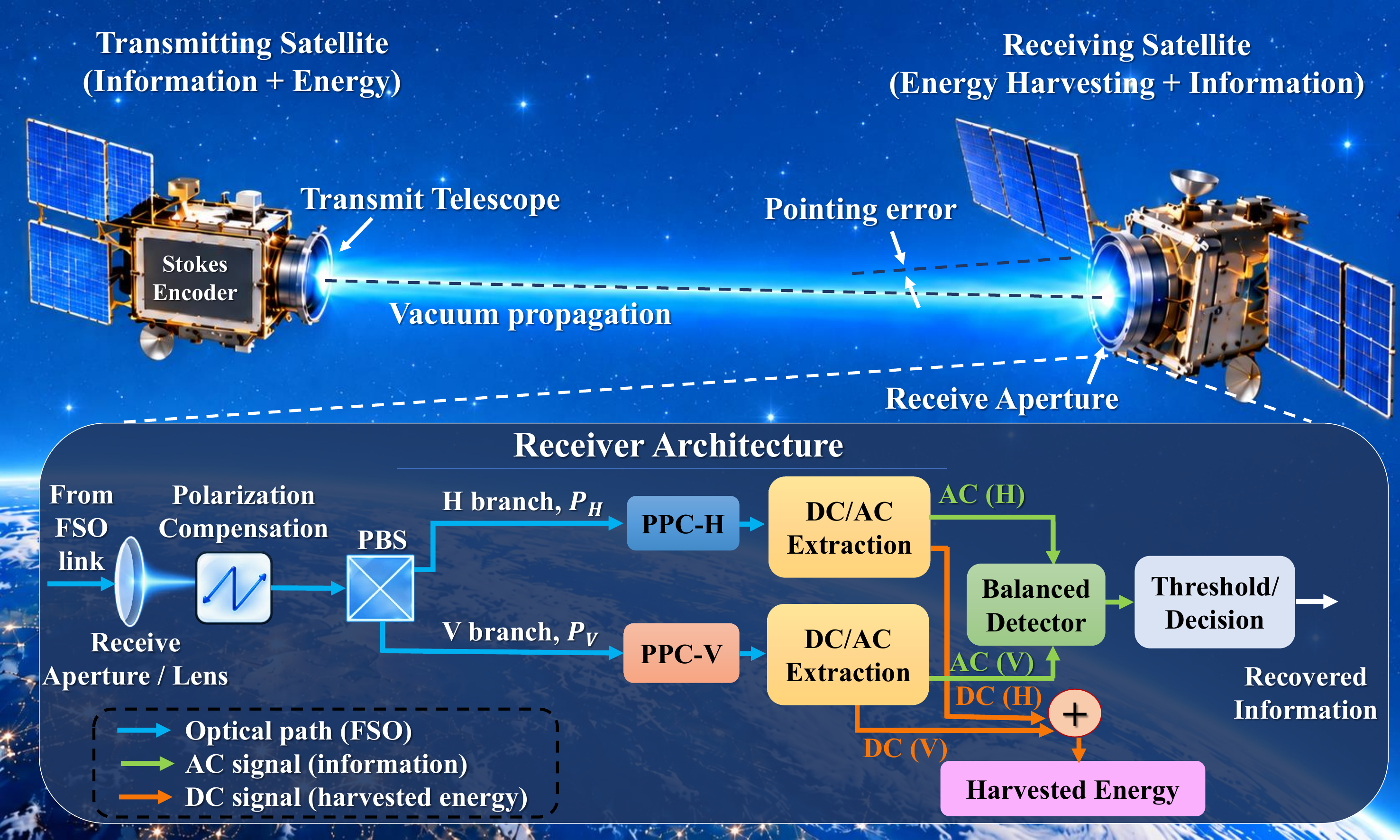}
\caption{Stokes-orthogonal inter-satellite FSO architecture with PBS-separated ac information recovery and dc photovoltaic energy harvesting.}
    \label{fig:fig1}
\end{figure}

\section{Vacuum Gaussian Channel and Pointing}

\subsection{Diffraction and Exact Circular-Aperture Reference}

For transmitter $1/e^2$ intensity radius $w_0$, Gaussian propagation gives~\cite{ref5}:
\begin{equation}
z_R
=
\frac{\pi w_0^2}{\lambda},
\qquad
w_L
=
w_0
\sqrt{
1+
\left(
\frac{L}{z_R}
\right)^2
}.
\label{eq:gaussian_propagation}
\end{equation}

where $w_L$ is the receiver-plane $1/e^2$ radius. Conditional on radial centroid displacement $r$, the normalized irradiance at receiver coordinate $\boldsymbol{\rho}$ is
\begin{equation}
i(\boldsymbol{\rho}\mid r)
=
\frac{2}{\pi w_L^2}
\exp
\left[
-\frac{2\left\lVert\boldsymbol{\rho}-\mathbf{r}\right\rVert^2}
{w_L^2}
\right],
\qquad
\int_{\mathbb{R}^2}
i(\boldsymbol{\rho}\mid r)\,
d\boldsymbol{\rho}
=
1.
\label{eq:normalized_irradiance}
\end{equation}

For a circular receive aperture of radius $a$, the exact collected fraction is
\begin{equation}
\begin{aligned}
h_{\mathrm{ex}}(r)
&=
\frac{4e^{-2r^2/w_L^2}}{w_L^2}
\int_{0}^{a}
\rho
e^{-2\rho^2/w_L^2}
I_0
\left(
\frac{4\rho r}{w_L^2}
\right)
d\rho
\\
&=
1-
Q_1
\left(
\frac{2r}{w_L},
\frac{2a}{w_L}
\right),
\end{aligned}
\label{eq:exact_aperture}
\end{equation}
where $I_0(\cdot)$ is the modified Bessel function and $Q_1(\cdot,\cdot)$ is the first-order Marcum $Q$-function~\cite{ref6}. Equation~\eqref{eq:exact_aperture}, or its Bessel-integral equivalent evaluated with a scaled $I_0$, is the numerical reference. It is bounded in $[0,1]$ and already includes diffraction-induced geometric loss; no separate inverse-square or aperture-area factor is multiplied into it.

\subsection{Pointing Law and Equivalent Beam}

Let $\theta_x,\theta_y \overset{\mathrm{iid}}{\sim}\mathcal{N}(0,\sigma_\theta^2)$. With $\sigma_s=L\sigma_\theta$, the radial displacement is Rayleigh \cite{Dabiri2026HierarchicalDL}:
\begin{equation}
r
=
L\sqrt{\theta_x^2+\theta_y^2},
\qquad
f_R(r)
=
\frac{r}{\sigma_s^2}
e^{-r^2/(2\sigma_s^2)},
\qquad
r\ge 0.
\label{eq:rayleigh_pointing}
\end{equation}

The exact aligned coupling, useful as a normalization check, is
\begin{equation}
h_{\mathrm{ex}}(0)
=
1-
\exp
\left(
-\frac{2a^2}{w_L^2}
\right).
\label{eq:aligned_coupling}
\end{equation}

For closed-form fading analysis, the standard equivalent-beam approximation is used~\cite{ref7}:
\begin{equation}
h(r)
\simeq
A_0
\exp
\left(
-\frac{2r^2}{w_{\mathrm{eq}}^2}
\right).
\label{eq:equivalent_beam}
\end{equation}

with
$
\nu
=
\frac{\sqrt{\pi}\,a}{\sqrt{2}\,w_L}$,
$
A_0
=
\operatorname{erf}^2(\nu)$,
$
w_{\mathrm{eq}}^2
=
w_L^2
\frac{\sqrt{\pi}\operatorname{erf}(\nu)}
{2\nu e^{-\nu^2}}$.
Here, $A_0$ is the aligned coupling parameter of the equivalent-beam formulation and is distinguished from the exact circular-aperture coupling in~\eqref{eq:aligned_coupling}. The numerical study evaluates both formulations under the same link conditions, enabling direct assessment of the analytical approximation against the exact circular-aperture reference.
Set
$
\xi
=
\frac{w_{\mathrm{eq}}}{2\sigma_s}$,
$
d
=
\xi^2$.
Transforming~\eqref{eq:rayleigh_pointing} through~\eqref{eq:equivalent_beam} gives
\begin{equation}
f_H(h)
=
\frac{d}{A_0^d}
h^{d-1},
\qquad
F_H(h)
=
\left(
\frac{h}{A_0}
\right)^d,
\qquad
0\le h\le A_0.
\label{eq:fading_pdf_cdf}
\end{equation}

\section{PBS Receiver and Electrical Noise}

\subsection{Complementary Branch Powers}

Let $\eta_t$, $\eta_r$, and $\eta_{\mathrm{pbs}}$ denote the terminal and common PBS insertion efficiencies, and define
$
P_0 = P_t \eta_t \eta_r \eta_{\mathrm{pbs}}$,
so that the received optical power after common losses and channel coupling is $P = P_0 h$.
Finite polarization contrast is parameterized by
$
c
=
\frac{1+\kappa}{2}$,
$
\ell
=
\frac{1-\kappa}{2}$,
and:
\begin{equation}
\mathrm{ER}
=
\frac{c}{\ell}
=
\frac{1+\kappa}{1-\kappa},
\qquad
0\le \kappa\le 1.
\label{eq:polarization_contrast}
\end{equation}

After the PBS, the H- and V-branch optical powers are
\begin{equation}
P_H(s,h)
=
\frac{P_0h}{2}(1+s\kappa),
\qquad
P_V(s,h)
=
\frac{P_0h}{2}(1-s\kappa).
\label{eq:branch_powers}
\end{equation}

Consequently,
\begin{equation}
P_H+P_V
=
P_0h,
\qquad
P_H-P_V
=
s\kappa P_0h.
\label{eq:sum_diff_powers}
\end{equation}

The first identity is optical power conservation after declared common losses; the second is the Stokes-$S_1$ observation attenuated only by contrast. No fraction $\rho P$ is diverted to an information detector: both branches are simultaneously photoconverting devices, their ac differential components carry data, and their dc outputs feed harvesting electronics. Bias-tee or diplexer circuits supporting simultaneous dc and ac extraction have been experimentally established~\cite{ref11}.
In the absence of electrical noise, normalized differencing gives
\begin{equation}
Z_0
=
\frac{I_H-I_V}{I_H+I_V}
=
s\kappa.
\label{eq:normalized_difference}
\end{equation}

Although $Z_0$ is useful for monitoring common optical fluctuations, a noisy ratio has a random denominator and is not automatically a sufficient or maximum-likelihood statistic. Performance is therefore derived for the physically direct balanced current
$
D
=
I_H-I_V$.
\subsection{Shot, Dark, Background, and Thermal Terms}

Let $R$ be the branch current responsivity. The sampled branch currents are
\begin{equation}
I_j = R P_j + n_j,
\qquad
j\in\{H,V\},
\label{eq:branch_currents}
\end{equation}
where $n_H$ and $n_V$ are independent zero-mean conditional Gaussian variables. The balanced output is
\begin{equation}
D
=
I_H-I_V
=
sR\kappa P_0 h+n_H-n_V.
\label{eq:balanced_output}
\end{equation}

For one-sided equivalent noise bandwidth $B$, background photocurrent per branch $I_{\mathrm{bg}}$, dark current $I_d$, load $R_L$, temperature $T$, and input-referred amplifier-current density $i_a$, take
\begin{equation}
\sigma_j^2
=
2qB
\left(
RP_j+I_{\mathrm{bg}}+I_d
\right)
+
\frac{4k_BTB}{R_L}
+
i_a^2 B.
\label{eq:branch_noise_variance}
\end{equation}

Using~\eqref{eq:sum_diff_powers}, the differential-noise variance is independent of the
transmitted symbol. Since the branch noises are independent,
$\sigma_D^2(h)=\sigma_H^2+\sigma_V^2$, and the resulting variance can be
written as
\begin{equation}
\sigma_D^2(h)=N_0+N_1h.
\label{eq:differential_noise_variance}
\end{equation}
Here, $N_0=4qB(I_{\mathrm{bg}}+I_d)+8k_BTB/R_L+2i_a^2B$
collects the background, dark-current, thermal, and amplifier-noise
contributions, while $N_1=2qBRP_0$ represents the signal-dependent
shot-noise coefficient.
The Gaussian shot-noise approximation requires sufficiently many detected photoelectrons over the decision interval. In the photon-starved limit, a joint Poisson model should replace~\eqref{eq:branch_noise_variance}; the thermal, dark, and background floor remains explicit here.

\section{Information-Transfer Performance}

\subsection{Detector Optimality and Conditional BER}

Define $A = R\kappa P_0$. The conditional electrical SNR is
\begin{equation}
\gamma(h)
=
\frac{A^2 h^2}{N_0 + N_1 h}.
\label{eq:conditional_snr}
\end{equation}

Then
\begin{equation}
D\mid(s,h)
\sim
\mathcal{N}
\left(
sAh,\,
N_0+N_1h
\right).
\label{eq:conditional_distribution}
\end{equation}

For known $h$, the equal-variance Gaussian hypotheses give
\begin{equation}
P_e(h)
=
Q
\left(
\frac{Ah}
{\sqrt{N_0+N_1h}}
\right),
\qquad
Q(x)
=
\frac{1}{2}
\operatorname{erfc}
\left(
\frac{x}{\sqrt{2}}
\right).
\label{eq:conditional_ber}
\end{equation}

This result preserves both the signal-dependent shot term and the nonzero noise floor.

\textit{Proposition 1 (Zero-threshold optimality):}
For equal priors, identical branches, and the noise model in~\eqref{eq:branch_noise_variance}, deciding
$
\hat{s}
=
\operatorname{sign}(D)
$
is maximum likelihood both with known $h$ and after marginalizing unknown $h$.

\textit{Proof:}
For fixed $h$, the two Gaussian hypotheses have means $\pm Ah$ and equal variance, hence cross at zero. Without channel knowledge, for any $x>0$, their marginal-density difference is
\begin{equation}
\begin{aligned}
p(x\mid +)-p(x\mid -)
&=
\int_{0}^{A_0}
\frac{
2e^{-\left(x^2+A^2h^2\right)/(2\sigma_D^2(h))}
}
{\sqrt{2\pi\sigma_D^2(h)}}
\\
&\quad\times
\sinh
\left(
\frac{xAh}{\sigma_D^2(h)}
\right)
f_H(h)\,dh
>0.
\end{aligned}
\label{eq:ml_proof_difference}
\end{equation}
Odd symmetry gives the reverse inequality for $x<0$. 

\subsection{Exact One-Dimensional Averages}

Under the equivalent-beam law, average BER is a single finite integral; under the exact aperture, Rayleigh pointing gives a second, independent one-dimensional reference integral:
\begin{equation}
\begin{aligned}
\overline{P}_e^{\mathrm{eb}}
&=
\frac{d}{A_0^d}
\int_{0}^{A_0}
Q
\left(
\frac{Ah}{\sqrt{N_0+N_1h}}
\right)
h^{d-1}\,dh,
\\
\overline{P}_e^{\mathrm{ex}}
&=
\int_{0}^{\infty}
P_e
\left(
h_{\mathrm{ex}}(r)
\right)
\frac{r}{\sigma_s^2}
e^{-r^2/(2\sigma_s^2)}
\,dr.
\end{aligned}
\label{eq:average_ber}
\end{equation}

Equation~\eqref{eq:average_ber} is the definitive performance expression. It is numerically benign after the substitutions
$
u
=
\left(
\frac{h}{A_0}
\right)^d
\in[0,1]
$
and
$
t
=
\frac{r^2}{2\sigma_s^2}$,
respectively. Gauss--Laguerre quadrature after the exponential-pointing transformation avoids absolute-tolerance loss in deep BER tails.

\subsection{Controlled Closed Forms}

If the signal-dependent shot term is uniformly small over the approximate support,
\begin{equation}
\epsilon_T
=
\frac{N_1A_0}{N_0}
\ll 1,
\qquad
P_e(h)
\simeq
Q(\alpha h),
\qquad
\alpha
=
\frac{A}{\sqrt{N_0}}.
\label{eq:thermal_dominant_condition}
\end{equation}

Integrating by parts with
$
x
=
\alpha A_0
$
yields
\begin{equation}
\overline{P}_{e,T}
=
Q(x)
+
\frac{2^{d/2-1}}
{\sqrt{\pi}\,x^d}
\gamma
\left(
\frac{d+1}{2},
\frac{x^2}{2}
\right),
\label{eq:closed_form_ber}
\end{equation}
where $\gamma(a,z)$ is the lower incomplete gamma function. This formula is exact for $N_1=0$ and otherwise controlled by~\eqref{eq:thermal_dominant_condition}.

Conversely, define the crossover
$
h_c
=
\frac{N_0}{N_1}$.
Since the pointing density always reaches $h=0$, shot-noise dominance cannot be uniform when $N_0>0$.

\begin{equation}
h_c
=
\frac{N_0}{N_1},
\qquad
F_H\!\left(\min\{h_c,A_0\}\right)
\ll 1.
\label{eq:shot_dominance_condition}
\end{equation}

Neglecting $N_0$ on the remaining probability mass gives
$
P_e(h)
\simeq
Q\!\left(\beta\sqrt{h}\right)$,
$
\beta
=
\frac{A}{\sqrt{N_1}}$.
For
$
y
=
\beta\sqrt{A_0},
$

\begin{equation}
\overline{P}_{e,S}
=
Q(y)
+
\frac{2^{d-1}}
{\sqrt{\pi}\,y^{2d}}
\gamma
\left(
d+\frac{1}{2},
\frac{y^2}{2}
\right).
\label{eq:shot_dominant_ber}
\end{equation}

Equation~\eqref{eq:shot_dominant_ber} is exact if $N_0=0$ and is otherwise an asymptotic approximation qualified by~\eqref{eq:shot_dominance_condition}. Neither closed form replaces~\eqref{eq:average_ber} outside its regime.

\subsection{SNR Outage}

For required post-detection SNR $\gamma_{\mathrm{th}}>0$, define
\begin{equation}
P_{\mathrm{out}}(\gamma_{\mathrm{th}})
=
\Pr
\left\{
\gamma(H)<\gamma_{\mathrm{th}}
\right\}.
\label{eq:snr_outage_def}
\end{equation}

For $A>0$, the mapping
\begin{equation}
\gamma(h)
=
\frac{A^2h^2}{N_0+N_1h},
\qquad
\gamma'(h)
=
\frac{
A^2h(2N_0+N_1h)
}{
(N_0+N_1h)^2
}
\ge 0
\label{eq:snr_monotonicity}
\end{equation}
is monotone. The positive inverse threshold is
\begin{equation}
h_{\mathrm{th}}
=
\frac{
\gamma_{\mathrm{th}}N_1
+
\sqrt{
(\gamma_{\mathrm{th}}N_1)^2
+
4A^2\gamma_{\mathrm{th}}N_0
}
}{
2A^2
}.
\label{eq:snr_inverse_threshold}
\end{equation}

Thus, under the equivalent beam,
\begin{equation}
P_{\mathrm{out}}
=
\begin{cases}
0,
&
h_{\mathrm{th}}\le 0,
\\[1mm]
\left(
\dfrac{h_{\mathrm{th}}}{A_0}
\right)^d,
&
0<h_{\mathrm{th}}<A_0,
\\[3mm]
1,
&
h_{\mathrm{th}}\ge A_0.
\end{cases}
\label{eq:snr_outage_closed_form}
\end{equation}

For $A=0$, including $\kappa=0$, outage equals one for every positive threshold. Exact-aperture outage is obtained by solving the monotone $h_{\mathrm{ex}}(r)$ crossing and using the Rayleigh tail; numerical inversion is preferable to treating~\eqref{eq:fading_pdf_cdf} as exact. 

\section{Nonlinear Photovoltaic Energy Analysis}

\subsection{Ideal Single-Diode MPP in Lambert-$W$ Form}

Each PBS output illuminates an identical $N_s$-junction PPC. We adopt the standard exponential multi-junction photovoltaic model, which provides a tractable nonlinear relation between incident optical power and the maximum extractable electrical power:

\begin{equation}
I(V;P)
=
I_{\mathrm{ph}}(P)
-
I_0
\left(
e^{V/a_T}-1
\right),
\label{eq:ppc_iv}
\end{equation}
where
\begin{equation}
a_T
=
N_s n_i
\frac{k_B T}{q},
\qquad
I_{\mathrm{ph}}(P)
=
\mathcal{R}P,
\qquad
\mathcal{R}
=
\eta_q
\frac{q\lambda}{N_s h_p c_0}.
\label{eq:ppc_parameters}
\end{equation}

Here $n_i$ is the effective ideality factor, $h_p$ is Planck's constant, $c_0$ is the speed of light, and $\eta_q$ is the effective photon-to-series-current yield. The factor $N_s$ in $\mathcal{R}$ represents equal photon/current allocation among matched stacked junctions; a measured spectral-response curve should replace it for a particular device.
Maximizing $V I(V;P)$ gives
\begin{equation}
I_{\mathrm{ph}}+I_0
=
I_0
e^{V/a_T}
\left(
1+\frac{V}{a_T}
\right).
\label{eq:mpp_stationarity}
\end{equation}

Set
$
C(P)
=
1+\frac{I_{\mathrm{ph}}(P)}{I_0}
$
and
$
W
=
W_0(eC)$.
The unique nonnegative maximum-power point (MPP) is
\begin{equation}
V_{\mathrm{mp}}
=
a_T(W-1),
\qquad
I_{\mathrm{mp}}
=
(I_{\mathrm{ph}}+I_0)
\left(
1-\frac{1}{W}
\right),
\label{eq:mpp_voltage_current}
\end{equation}
and the harvested-power transfer function is
\begin{equation}
g(P)
=
a_T
(I_{\mathrm{ph}}+I_0)
\frac{(W-1)^2}{W},
\qquad
g(0)=0.
\label{eq:harvested_power_function}
\end{equation}

The principal real Lambert function $W_0$ is sufficient because $eC \ge e$, yielding a unique closed-form MPP through the standard Lambert-$W$ treatment of exponential equations~\cite{ref14}. Near-infrared InGaAs/InP and InGaAsP converters provide experimentally relevant efficiency and simultaneous-reception benchmarks~\cite{ref11,ref12}. The saturation-current parameter $I_0$ is calibrated to a measured MPP operating point, thereby anchoring the analytical converter model to representative near-infrared PPC performance.

\subsection{Symbol-Invariant Nonlinear Harvesting}

For branch input $P=P_0h$, the quasi-static harvested output for symbol $s$ is
\begin{equation}
G_s(P)
=
g\left(
\frac{P}{2}(1+s\kappa)
\right)
+
g\left(
\frac{P}{2}(1-s\kappa)
\right).
\label{eq:symbol_harvested_power}
\end{equation}

\textit{Proposition 2 (Nonlinear energy neutrality):}
If the two branch converters and their MPP controllers have the same transfer function $g(\cdot)$, then
$
G_{+1}(P)=G_{-1}(P)
$
for every $P\ge 0$, every $\kappa\in[0,1]$, and any nonlinear $g$.

\textit{Proof:}
Using~\eqref{eq:polarization_contrast},
$
G_{+1}(P)
=
g(cP)+g(\ell P)
=
g(\ell P)+g(cP)
=
G_{-1}(P)$.
The symbol changes branch labels, not the unordered pair of optical inputs.
This result is stronger than a linear-efficiency argument: no Jensen or small-signal approximation is used. It does not imply that absolute harvested power is independent of $\kappa$; because $g$ is nonlinear,
$
g(cP)+g(\ell P)
$
generally changes with contrast.
The exact and equivalent-beam average harvested powers are
\begin{equation}
\begin{aligned}
\overline{P}_{\mathrm{EH}}^{\mathrm{ex}}
&=
\int_{0}^{\infty}
\left[
g\!\left(cP_0h_{\mathrm{ex}}(r)\right)
+
g\!\left(\ell P_0h_{\mathrm{ex}}(r)\right)
\right]
f_R(r)\,dr,
\\
\overline{P}_{\mathrm{EH}}^{\mathrm{eb}}
&=
\frac{d}{A_0^d}
\int_{0}^{A_0}
\left[
g(cP_0h)+g(\ell P_0h)
\right]
h^{d-1}\,dh.
\end{aligned}
\label{eq:average_harvested_power}
\end{equation}

The framework naturally supports two energy-extraction timescale regimes. When the branch controller follows the instantaneous polarization-dependent optical level, the harvested power is described by~\eqref{eq:symbol_harvested_power}. For high-rate signaling relative to the electrical energy-extraction dynamics, an equiprobable balanced stream presents an average optical input $P/2$ to each branch, yielding
\begin{equation}
G_{\mathrm{fast}}(P)
=
2g\left(\frac{P}{2}\right).
\label{eq:fast_symbol_harvesting}
\end{equation}

These two expressions characterize the quasi-static and fast-signaling operating regimes, respectively, while preserving symbol-invariant energy harvesting in both cases. Together they connect the proposed architecture to a broad range of PPC and MPP-controller bandwidths~\cite{ref9,ref11}.
If branches are unequal, with transfer functions $g_H$ and $g_V$, the residual symbol dependence is exactly
\begin{equation}
\begin{aligned}
\Delta G(P)
&=
G_{+1}-G_{-1}
\\
&=
\left[
g_H(cP)-g_H(\ell P)
\right]
-
\left[
g_V(cP)-g_V(\ell P)
\right].
\end{aligned}
\label{eq:branch_mismatch_deltaG}
\end{equation}

Equation~\eqref{eq:branch_mismatch_deltaG} supplies a direct calibration target: equality of the incremental responses at $cP$ and $\ell P$ is sufficient for neutrality.

\subsection{Consistency and Limiting Cases}

The formulation obeys the following dimensional and physical checks. As $P_t\rightarrow 0$, $A,N_1\rightarrow 0$, so $P_e(h)\rightarrow 1/2$ and $g(P)\rightarrow 0$. As $\kappa\rightarrow 1$, $(c,\ell)\rightarrow(1,0)$, $G_s(P)=g(P)$, and separation is ideal; as $\kappa\rightarrow 0$, the balanced mean vanishes and BER approaches $1/2$, while both branches receive $P/2$. As $\sigma_\theta\rightarrow 0$, pointing collapses to the aligned coupling, and $a\rightarrow\infty$ gives $h_{\mathrm{ex}}\rightarrow 1$ for finite $r$. 
Moreover, $h$, $A_0$, $c$, $\ell$, $\kappa$, and $d$ are dimensionless; $P_0$, $P_j$, and $g$ are in watts; $R$ is in amperes per watt; $N_0$ and $N_1h$ are in amperes squared; $Ah$ is in amperes; and $\gamma$ is dimensionless.

\section{Numerical Results and Simulation Setup}

The numerical study adopts $\lambda=1530~\mathrm{nm}$, an inter-satellite range $L=100~\mathrm{km}$, transmitted optical power $P_t=2~\mathrm{W}$, transmitter beam radius $w_0=25~\mathrm{mm}$, receive-aperture radius $a=100~\mathrm{mm}$, per-axis pointing jitter $\sigma_\theta=2~\mu\mathrm{rad}$, and polarization extinction ratio $\mathrm{ER}=20~\mathrm{dB}$. The near-infrared wavelength, link scale, compact apertures, and watt-class optical source are representative of published small-spacecraft terminal studies~\cite{ref2,ref3,ref4}; the particular $w_0$, $\sigma_\theta$, insertion efficiencies $(\eta_t,\eta_r,\eta_{\mathrm{pbs}})=(0.80,0.80,0.95)$, and extinction ratio are modeling assumptions selected for investigation rather than measured performance claims. The receiver uses $B=1~\mathrm{GHz}$, $T=293.15~\mathrm{K}$, $R_L=25~\Omega$, and $I_d=30~\mathrm{nA}$ per branch, with background and amplifier noise retained parametrically. For the four-junction PPC benchmark, $N_s=4$, $n_i=1.05$, and $\eta_q=0.90$ are used, while $I_0$ is calibrated only at a stated near-infrared MPP operating point drawn from the multi-junction measurements in~\cite{ref11}; reported sweeps will distinguish these declared model choices from literature-supported hardware ranges and will compare~\eqref{eq:exact_aperture} against~\eqref{eq:equivalent_beam} without inserting an additional geometric-loss factor.

Fig.~2 illustrates the average BER as a function of launched optical power for three levels of residual pointing jitter. At low transmit powers, all three cases remain in a high-BER regime, whereas increasing $P_t$ produces a pronounced transition to rapidly decreasing BER. The onset and rate of this transition depend strongly on $\sigma_{\theta}$: the $6~\mu\mathrm{rad}$ case enters the low-BER region first and exhibits the steepest decay, while the $8$ and $10~\mu\mathrm{rad}$ cases are progressively shifted toward higher required transmit powers. For example, a BER near $10^{-2}$ is reached at approximately $13$, $17$, and $23~\mathrm{dBm}$ for $\sigma_{\theta}=6$, $8$, and $10~\mu\mathrm{rad}$, respectively, demonstrating the increasing power penalty associated with stronger pointing fluctuations. The Monte Carlo markers closely follow the analytical curves throughout the resolved BER range, while the closed-form traces remain visually coincident with the corresponding analytical results. Thus, the figure establishes both the strong sensitivity of communication reliability to residual pointing jitter and the close agreement among the analytical, closed-form, and simulation-based BER evaluations.

\begin{figure}[t]
    \centering
    \includegraphics[width=\columnwidth]{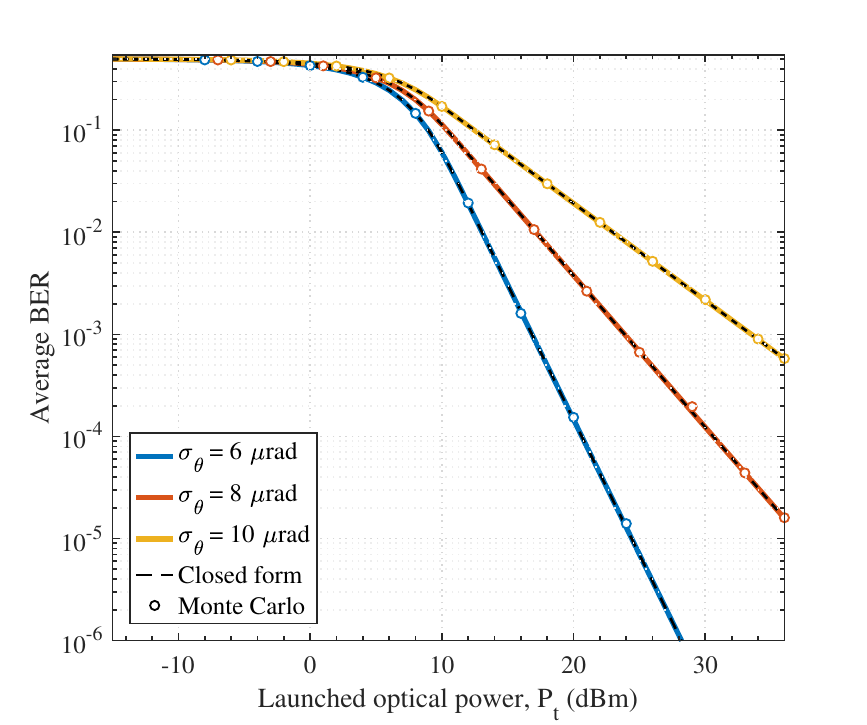}
\caption{Average BER versus launched optical power for per-axis pointing jitter $\sigma_{\theta}=6$, $8$, and $10~\mu\mathrm{rad}$.}
    \label{fig:fig2}
\end{figure}

\begin{figure}[t]
    \centering
    \includegraphics[width=\columnwidth]{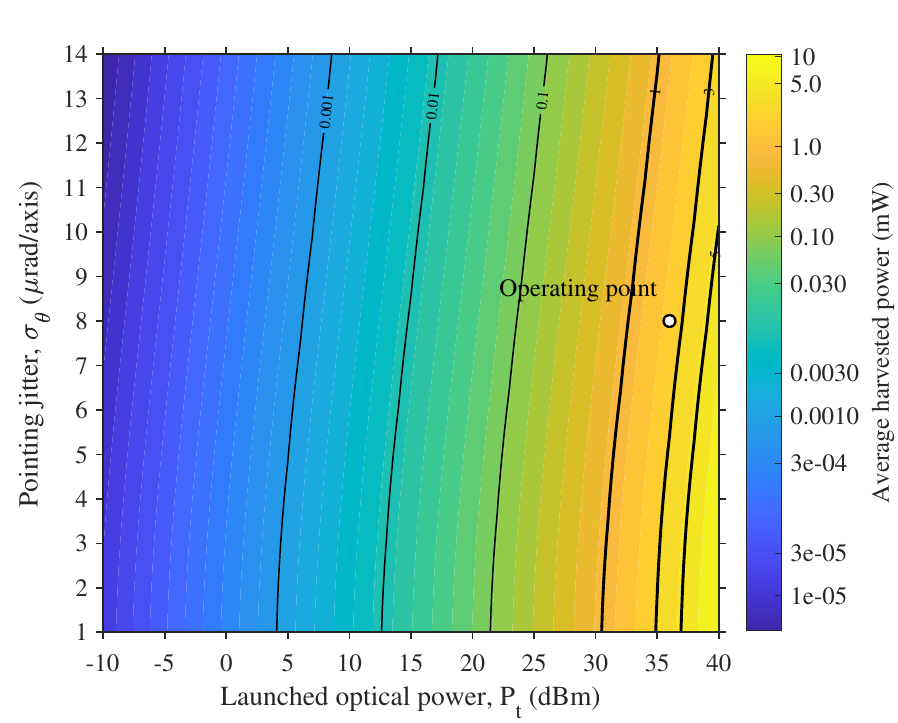}
\caption{Average harvested power versus launched optical power and per-axis pointing jitter, with labeled iso-power contours and the marked operating point.}
    \label{fig:fig3}
\end{figure}

Fig.~3 presents the harvested-power design map of the proposed architecture in the $(P_t,\sigma_{\theta})$ plane and shows that the average harvested power increases monotonically with launched optical power while decreasing as the pointing jitter grows. The color transition and contour progression indicate that $P_t$ is the dominant control variable, whereas $\sigma_{\theta}$ introduces a clear penalty by shifting a given harvested-power level toward higher required transmit powers. This behavior is reflected in the contour geometry: the iso-power lines are nearly vertical but tilt to the right as $\sigma_{\theta}$ increases, meaning that stronger pointing fluctuations must be compensated by additional launched power to preserve the same energy level. For example, at $P_t=30~\mathrm{dBm}$, the harvested power drops from about $0.870~\mathrm{mW}$ at $\sigma_{\theta}=1~\mu\mathrm{rad/axis}$ to about $0.271~\mathrm{mW}$ at $14~\mu\mathrm{rad/axis}$, while at $P_t=40~\mathrm{dBm}$ it decreases from about $10.73$ to $3.36~\mathrm{mW}$ over the same jitter range. The marked operating point at $P_t=36~\mathrm{dBm}$ and $\sigma_{\theta}=8~\mu\mathrm{rad/axis}$ lies between the $1$ and $3~\mathrm{mW}$ contours and corresponds to approximately $2.31~\mathrm{mW}$, which is consistent with the surrounding map structure. Overall, the figure provides a compact design-oriented visualization of the transmit-power and pointing-stability tradeoff that governs the achievable harvested energy in the considered inter-satellite FSO link.

Fig.~4 compares the BER--harvested-power relationship of the proposed architecture with that of the ideal power-splitting benchmark at a fixed launched optical power. The dashed benchmark curves exhibit the expected tradeoff: as the operating point moves rightward toward larger harvested power, the BER increases monotonically, so improved energy extraction is obtained at the cost of reduced communication reliability. The three benchmark curves are also ordered by pointing jitter, with $\sigma_{\theta}=6~\mu\mathrm{rad}$ occupying the most favorable lower-right region and $\sigma_{\theta}=10~\mu\mathrm{rad}$ shifting upward and leftward, indicating that stronger pointing fluctuations simultaneously worsen BER and reduce the achievable harvested power. In contrast, the proposed operating points are displaced to the right of the corresponding benchmark trajectories while remaining at substantially lower BER. For example, at $\sigma_{\theta}=8~\mu\mathrm{rad}$, the proposed point is located at about $2.31~\mathrm{mW}$ with BER $1.6\times10^{-5}$, whereas the benchmark has already risen to about $1.7\times10^{-4}$ near $1.81~\mathrm{mW}$. Similarly, for $\sigma_{\theta}=10~\mu\mathrm{rad}$, the proposed point reaches about $1.87~\mathrm{mW}$ at BER $5.8\times10^{-4}$, while the benchmark is around $2.7\times10^{-3}$ near $1.46~\mathrm{mW}$. For $\sigma_{\theta}=6~\mu\mathrm{rad}$, the proposed point attains the largest harvested power, about $2.84~\mathrm{mW}$, with BER below the displayed $10^{-8}$ floor. Thus, the figure shows that the proposed scheme reaches the high-harvested-power end of the benchmark operating range without following its associated BER degradation.

\begin{figure}[t]
    \centering
    \includegraphics[width=\columnwidth]{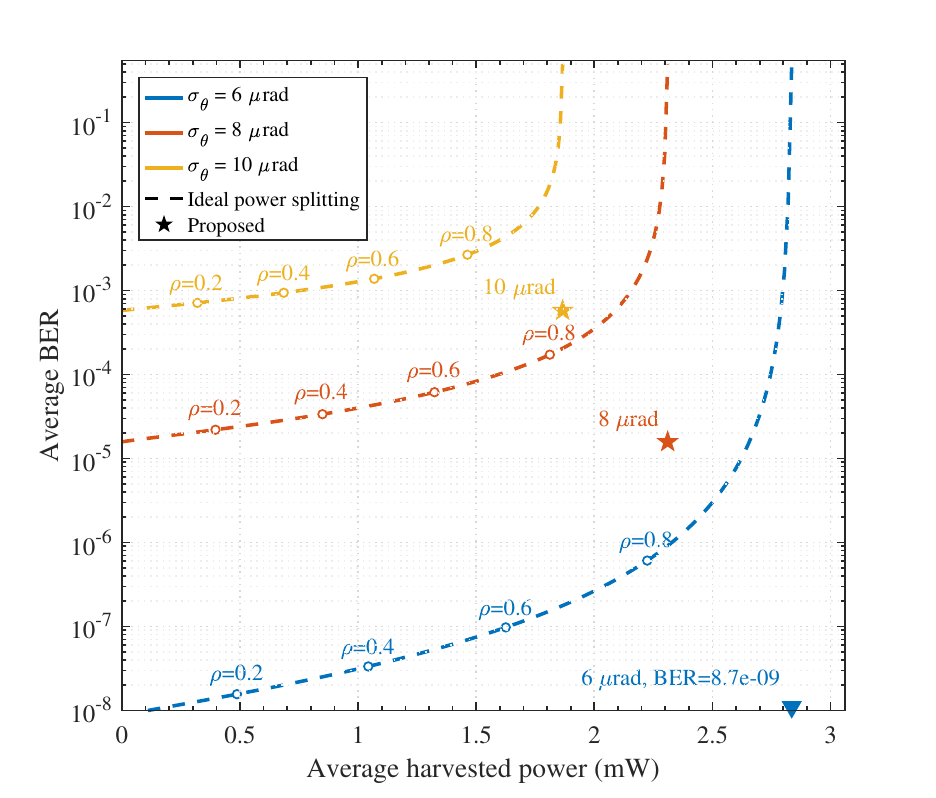}
\caption{Average BER versus average harvested power at $P_t=36~\mathrm{dBm}$ for the proposed architecture and the ideal power-splitting benchmark, for $\sigma_{\theta}=6$, $8$, and $10~\mu\mathrm{rad}$.}
    \label{fig:fig4}
\end{figure}

\section{Conclusion}

A constant-total-power Stokes-orthogonal architecture was developed for joint information transfer and optical energy harvesting over inter-satellite FSO links. The study established that polarization-domain signaling enables the information-bearing differential component and the energy-bearing dc component to be extracted from the same received optical power without introducing an explicit information–energy allocation factor. For matched nonlinear conversion branches, the transmitted symbols exchange the branch powers while preserving their unordered pair, which guarantees symbol-invariant harvested energy beyond a linear-efficiency model. The numerical results further showed that residual pointing jitter governs both communication reliability and harvested power, while the analytical BER expressions closely agree with Monte Carlo evaluation across the investigated operating range. Most importantly, comparison with ideal lossless power splitting showed that the proposed architecture can retain high harvested power without incurring the corresponding BER degradation associated with conventional resource allocation. These results provide a compact analytical and design reference for power-neutral joint information and energy transfer in pointing-impaired inter-satellite optical links.

\section*{Acknowledgment}
This work was supported by the Qatar Research Development and Innovation Council (QRDI) under Grant No. NPRP14C-0909-210008 and by research funding from Hamad Bin Khalifa University under the Thematic Research Grant Program Cycle 3. The statements made herein are solely the responsibility of the authors. The content is solely the responsibility of the authors and does not necessarily represent the official views of QRDI.

\bibliographystyle{IEEEtran}
\bibliography{myref}

\end{document}